\documentclass[a4paper,dvips,12pt]{article}
\usepackage{amsmath,amssymb,exscale}  
\usepackage{array,multicol}
\usepackage{afterpage,float,flafter}   
\usepackage{epsfig,rotating,pifont,fancybox}   
\makeatletter   
\@addtoreset{equation}{section}   
\makeatother   
   
\newcommand{\bea}{\begin{eqnarray}}   
\newcommand{\eea}{\end{eqnarray}}   
   
\newcommand{\NPB}[3]{\emph{ Nucl.~Phys.} \textbf{B#1} (19#2) #3}   
\newcommand{\PLB}[3]{\emph{ Phys.~Lett.} \textbf{B#1} (19#2) #3}   
\newcommand{\PRD}[3]{\emph{ Phys.~Rev.} \textbf{D#1} (19#2) #3}   
\newcommand{\PRL}[3]{\emph{ Phys.~Rev.~Lett.} \textbf{#1} (19#2) #3}

\newcommand{\EPJC}[3]{\emph{ Eur.~Phys.~J.} \textbf{C#1} (19#2) #3}

\title{   
\vspace*{-0.8cm}   
\begin{flushright}   
\normalsize{      
IEM-FT-189/99\\
IFT-UAM/CSIC-99-9\\   
\textsf{hep-ph/9903400}}\\ 
\end{flushright}    
\vspace{1cm}
\Large\textbf{Strong coupling unification and extra dimensions~\footnote{Work 
supported in part by CICYT, Spain, under contract AEN98-0816.}}
\vspace*{.5cm}
\author{\large
{A.~Delgado and M.~Quir{\'o}s}\\ \\
\emph{Instituto de Estructura de la Materia (CSIC), Serrano 123,}\\
\emph{E-28006-Madrid, Spain.}}}
\date{}   
\begin{document}
\maketitle

\vspace*{2cm}

\begin{abstract}
We analyze the implications of electroweak and strong coupling unification
in a very general class of models extending the minimal
supersymmetric standard model in $4+p$ dimensions 
$(p\geq 0)$. In general, electroweak precision data require the presence of
large extra dimensions (low compactification scales, $M_c$) 
and/or low unification scale, $M_U$.
In particular, the actual experimental value of the strong 
coupling at $M_Z$ imposes an upper bound on the compactification and 
unification scales. In four dimensional theories 
($M_c\simeq M_U$) with canonical hypercharge assignment we find 
$M_U\lesssim 10^{9}$ GeV. In theories with extra dimensions
($M_c<M_U$) we find $M_c\lesssim 10^7$ GeV, for a supersymmetric
spectrum at the TeV scale. 
\end{abstract}
\vspace{5.cm}   
   
\begin{flushleft}   
March 1999 \\   
\end{flushleft}   
\newpage

\section{Introduction} 
\label{introduction}
Gauge coupling unification is considered as a very suggestive hint of 
physics beyond the Standard Model (SM), since almost any 
fundamental theory describing physics at high scales predicts a sort of 
grand (or string) unification of gauge couplings. This idea led to rule
out the SM with gauge couplings unifying in a non-supersymmetric 
$SU(5)$~\cite{SMunif}, as the effective theory of electroweak and strong 
interactions, and
to favour its minimal supersymmetric extension (MSSM)~\cite{MSSMunif}, 
on the basis of LEP and
low energy precision measurements~\cite{PDG}. This is the most compelling 
`experimental'
indication in favour of supersymmetry at low scales (phenomenological 
supersymmetry). It predicts a big desert between the weak and unification
scales, $M_U\simeq 2\times 10^{16}$ GeV, populated by the MSSM and with gauge
coupling unifying at $M_U$ with a value $\alpha_U\simeq 1/24$.

However, the above picture where the MSSM unifies at $M_U$ is being 
jeopardized by the increasing precision of experimental data that demands 
increasing accurateness in gauge coupling unification. In particular, by
imposing unification of $SU(2)\times U(1)$ gauge couplings, and the
experimental input~\cite{PDG} \footnote{$\widehat{\alpha}_i(\mu)$ means 
the corresponding gauge coupling in the modified minimal subtraction 
$(\overline{\rm MS})$ renormalization scheme at the scale $\mu$.},
\begin{eqnarray}
\label{eweak}
\widehat{\alpha}_1^{-1}(M_Z)&=&58.98\pm 0.04\nonumber\\
\widehat{\alpha}_2^{-1}(M_Z)&=&29.57\pm 0.03\nonumber\\ 
M_Z&=&91.197\pm 0.007
\end{eqnarray}
it has been shown, using the two-loop renormalization group 
equations (RGE)~\cite{twoloop},
the $\overline{\rm MS}\rightarrow\overline{\rm DR}$~\cite{DR} conversion 
factors~\cite{MStoDR} and
the low energy supersymmetric threshold for minimal fine tuning,
that $\alpha_3(M_Z)\simeq 0.13$~\cite{LP}-\cite{pierce} 
which is $\sim$ five 
standard deviations away from the experimental value~\cite{PDG}
\begin{equation}
\label{strong}
\widehat{\alpha}_3(M_Z)=0.119\pm 0.002
\end{equation}
These results correspond to the case of an $R$-parity conserving MSSM. 
The case of
the MSSM with $R$-parity breaking couplings $\lambda$ has been recently 
analyzed in
Ref.~\cite{herbie}, where it is shown that the predicted value of 
$\alpha_3(M_Z)$ is very insensitive to the values of $\lambda(M_U)$ 
except possibly in the region close or beyond the non-perturbative 
values $\lambda^2(M_U)\gtrsim 4\pi$, where the prediction can be brought to
within $\sim 2\sigma$ of the observed value.

This implies that, in order to achieve successful unification, one has to 
rely on unknown high-energy thresholds. Without having any 
information about the high-energy theory this solution is unlikely to be
experimentally confirmed. We will call this problem the strong 
unification problem.

In this paper we propose another solution to the strong unification 
problem based on the possible existence of large extra dimension(s). The
possibility of large extra dimensions 
(as large as at the TeV$^{-1}$ length)~\cite{ignatios}
feeling the gauge interactions has been recently shown as very
natural in a large class of string theories, in particular in 
type I~\cite{lykken}-\cite{luis} 
and type IIB~\cite{pioline} string vacua. 
Unlike in the perturbative heterotic string case, where one was led to 
identify the compactification scale, $M_c$, with the unification, $M_U$,
and the string, $M_{st}$, 
scales $\sim 10^{18}$ GeV (creating a problem of around two orders of magnitude
in the prediction of the Newton constant), in type I and IIB strings or in the 
non-perturbative regime of the heterotic string~\cite{nonpert}, 
the prediction of 
the string/unification scale is relaxed and can be as low as the TeV 
scale.
This fortunate accident allows to relax the condition of a large unification
scale $M_U$ and try to find unifying models where the strong coupling
problem is resolved without any dependence on features of the underlying
high-energy theory. This is the issue that will be dealt with in this paper.
We will make a systematic study of models where gauge couplings
unify such that the strong coupling at $M_Z$ lies inside the experimental 
range of Eq.~(\ref{strong}).

In section \ref{logarithmic} we will study 4D models that unify at 
any value of $M_c\sim M_U$
below the Planck scale and that solve the strong unification problem. 
Interestingly enough we will find that all satisfactory models unify at scales
$\lesssim 10^{11}$ GeV. In section 3 we will open up the possibility of
unification in $(4+p)$ dimensions provided that $M_c<M_U$. In
these models the evolution of gauge couplings is logarithmic
between the weak scale and $M_c$ and power-law between $M_c$ and $M_U$.
We will find the general class of models that unify at one-loop in 
the same way as the MSSM. However by 
including two-loop corrections in the 4D theory, below $M_c$, we find that
the experimental range (\ref{strong}) implies that $M_U\lesssim 10^9$ GeV.
On the basis of the previous results we claim that gauge coupling
unification prefer the presence of large extra dimension(s) feeling the gauge
interactions. Finally section \ref{conclusions} contains our conclusions.

\section{Logarithmic unification}
\label{logarithmic}
We will start this section with some general considerations
based on the one-loop unification of gauge couplings:
\begin{equation}
\alpha_i^{-1}(\Lambda)=\alpha_i^{-1}(\mu)-\frac{b_i}{2\pi}\log\frac{\Lambda}
{\mu}
\label{alpha1}
\end{equation}
where $b_i\, (i=1,2,3)$ are the one-loop beta-functions of the model and we
are (GUT) normalizing the hypercharge as: $\alpha_1=k_1
g^{\prime\,2}/4\pi$, where $g^\prime$ is the $U(1)_Y$ gauge coupling and
$k_1=5/3$~\footnote{We will relax this condition in section
\ref{conclusions}.}. The unification condition is then
$\alpha_1(M_U)=\alpha_2(M_U)= \alpha_3(M_U)$. 

The couplings $\alpha_1$ and $\alpha_2$ meet at one-loop at the scale:
\begin{equation}
\label{MU1}
M_U^{(1)}=\mu \,e^{\,\alpha_{12}^{-1}(\mu)\,\frac{2\pi}{b_1-b_2}}
\end{equation}
with a value
\begin{equation}
\label{aU1}
\alpha_U^{(1)\, -1}=\alpha_1^{-1}(\mu)-\alpha_{12}^{-1}(\mu)\ 
\frac{b_1}{b_1-b_2}=\alpha_2^{-1}(\mu)-\alpha_{12}^{-1}(\mu)\ 
\frac{b_2}{b_1-b_2}
\end{equation}
where we are using the notation 
$\alpha_{ij}^{-1} \equiv\alpha_i^{-1}-\alpha_j^{-1}$. The one
loop prediction of the strong coupling is then:
\begin{equation}
\label{alphas1}
\alpha_3^{(1)\, -1}(\mu)=\alpha_2^{-1}(\mu)-\alpha_{12}^{-1}(\mu)\ 
\frac{b_2-b_3}{b_1-b_2}=
\alpha_1^{-1}(\mu)-\alpha_{12}^{-1}(\mu)\ \frac{b_1-b_3}{b_1-b_2}\
.
\end{equation}

Of course Eqs.~(\ref{MU1}), (\ref{aU1}) and (\ref{alphas1}) should be
compared with the experimental data (\ref{eweak}) and (\ref{strong}) at
the scale $\mu=M_Z$.
From Eqs.~(\ref{MU1}), (\ref{aU1}) and (\ref{alphas1}) it is
obvious that given a unifying theory with beta-coefficients
$b_i$, the new theory with beta-coefficients $\beta_i=b_i+c$,
where $c$ is a constant, unifies with the same value of
$M_U^{(1)}$ and $\alpha_3^{(1)}(M_Z)$ but different value of
$\alpha_U^{(1)}$. This happens for instance in the case where we
add to the initial theory complete representations of a grand
unification group which contains the gauge group of the theory.
For instance in the case of the MSSM it happens when we add
complete representations of $SU(5)$, or any other group
containing $SU(5)$, e.g. $SO(10)$ or $E_6$. However this class
of models, where $M_U^{(1)}$ is unchanged, is too restrictive
and we are not interested in it. Instead we will focus on
the more general class of models, characterized by the beta-coefficients
$\beta_i$, such that only $\alpha_3^{(1)}(M_Z)$ is unchanged.
Using (\ref{alphas1}) we can see that these models must satisfy
the equations:
\begin{equation}
\label{ecuaciones}
\frac{b_1-b_3}{b_1-b_2}=\frac{\beta_1-\beta_3}{\beta_1-\beta_2},\quad
\frac{b_2-b_3}{b_1-b_2}=\frac{\beta_2-\beta_3}{\beta_1-\beta_2}
\end{equation}

The general solution to (\ref{ecuaciones}) is given by
\begin{equation}
\label{class}
\epsilon^{ijk}\left(b_i-b_j\right)\beta_k=0
\end{equation}
It is obvious that $\beta_i=b_i$ itself belongs to the class
defined by (\ref{class}), as well as $\beta_i=b_i+c$. However,
there are other models for which the prediction of
$\alpha_3^{(1)}(M_Z)$ is the same but with different values of
$M_U^{(1)}$ and $\alpha_U^{(1)}$. 

As an example, in the MSSM the one-loop beta-coefficients are given by
\begin{equation}
\label{betaMSSM}
b_i=(33/5,1,-3)
\end{equation}
which lead, from (\ref{alphas1}), to 
$\alpha_3^{(1)}(M_Z)=0.117$, and from Eqs.~(\ref{MU1}) and
(\ref{aU1}), to $M_U^{(1)}=2\times 10^{16}$ GeV and
$\alpha_U^{(1)}=1/24.3$. The class of models which predict
the same value of $\alpha_3^{(1)}(M_Z)$ corresponds to
beta-coefficients satisfying the equation
\begin{equation}
\label{classMSSM}
5\,\beta_1-12\,\beta_2 +7\,\beta_3=0
\end{equation}
Then the different models satisfying Eq.~(\ref{classMSSM}) unify
at different values of $M_U^{(1)}$ and nevertheless they lead to
$\alpha_3^{(1)}(M_Z)=0.117$. Some of these models will be discussed
later on in this section in our general search (see Tables~\ref{tab:cano}
and~\ref{tab:non-cano}). 

Two-loop corrections can
lead to little differences in the final value of $\alpha_3(M_Z)$
as we will see later on in this section, where we will identify
several models which belong to the same class.
However two loop corrections have been found to be very stable
inside the same class, since there is a compensating effect 
for models that unify at smaller values of $M_U$ due to the fact
that they have more extra matter, and the class (\ref{class})
leads approximatively to the same value of $\alpha_3(M_Z)$.

\subsection{The MSSM}

The MSSM belongs to the class (\ref{classMSSM}) and leads, as we
have already stated, to $\alpha_3^{(1)}(M_Z)=0.117$. We will
review the calculation of $\alpha_3(M_Z)$ including two-loop
corrections.

The RGE predictions for the gauge couplings in the MSSM can be
written assuming unification as:
\begin{equation}
\label{RGE}
\alpha_i^{-1}(M_Z)=\alpha_U^{-1}+\frac{b_i}{2\pi}\log\frac{M_U}{M_Z}+
\frac{1}{4\pi}\sum_{j=1}^{3}\frac{b_{ij}}{b_j}\log\frac{\alpha_j(M_U)}
{\alpha_j(M_Z)}+\Delta_i
\end{equation}
where the two-loop beta coefficients matrix is given by:
\begin{equation}
\label{matriz}
b_{ij}=\left(
\begin{array}{ccc} 
199/25 & 27/5 & 88/5 \\
9/5 & 25 & 24 \\
11/5 & 9 & 14
\end{array}
\right)
\end{equation}
and $\Delta_i$ contains the $\overline{\rm MS}$ to
$\overline{\rm DR}$ renormalization scheme conversion factors
($C_2(G_i)/12\pi$) and the low-energy supersymmetric thresholds.
Assuming for the latter that gaugino masses are degenerate at
the unification scale one gets, keeping the dominant contributions to
the thresholds~\footnote{Other contributions to $\Delta_i$ as
top and Higgs thresholds and the contribution from the Yukawa
couplings are negligible. A systematic analysis of all these
effects can be found in Ref.~\cite{LP}.}~\cite{LP,pierce}:
\begin{eqnarray}
\label{Deltas}
\Delta_2&=&\frac{1}{6\pi}\nonumber\\
\Delta_3&=&\frac{1}{4\pi}+\frac{1}{28\pi}\left(28\log\frac{\alpha_2(M_Z)}
{\alpha_3(M_Z)}+19\log\frac{M_{\rm SUSY}}{M_Z}\right)
\end{eqnarray}
where $M_{\rm SUSY}$ denotes a common supersymmetry mass.

The unification prediction from (\ref{RGE}) is 
\begin{eqnarray}
\label{predMSSM}
M_U^{\rm MSSM}&=&3.7\times 10^{16}\, {\rm GeV}\nonumber\\
\alpha_U^{\rm MSSM}&=&1/22.9\nonumber \\
\alpha_3^{\rm MSSM}(M_Z)&=&0.131 \quad .
\end{eqnarray}
Therefore we see that the prediction of the MSSM is $\sim$ 5--6
standard deviations away from the experimental value
(\ref{strong}) and also that the amount of all two-loop
corrections in $\alpha_3(M_Z)$ can be quantified to be
\begin{equation}
\label{inc2}
\Delta_{\rm two-loop\ }\alpha_3(M_Z)\simeq 0.014
\end{equation}
This quantity is the clue to find models solving the strong
unification problem: i.e. models predicting
\begin{equation}
\label{desired}
\alpha_3^{(1)}(M_Z)\simeq 0.105
\end{equation}
which will be done next.

\subsection{General analysis}
The first step towards the search of models which can solve the 
strong unification problem is to analyze the one-loop equations (\ref{MU1}),
(\ref{aU1}) and (\ref{alphas1}) for a general model defined by general one-loop
beta-coefficients $\beta_i\equiv b_i+\Delta b_i$ where $b_i$ are
the MSSM beta-coefficients, Eq.~(\ref{betaMSSM}). 

\begin{figure}[ht]
\centering
\epsfig{file=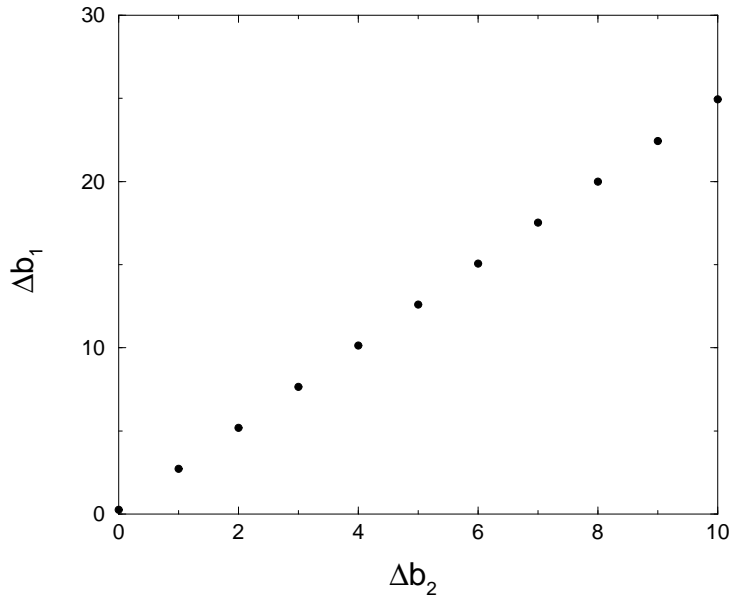,width=0.6\linewidth}
\caption{Plot of $\Delta b_1$ as a function of $\Delta b_2$ for $\Delta b_3=0$
and $\alpha_3^{(1)}(M_Z)\simeq 0.105$.}     
\label{fig:b1-b2}
\end{figure}

\begin{figure}[ht]
\centering
\epsfig{file=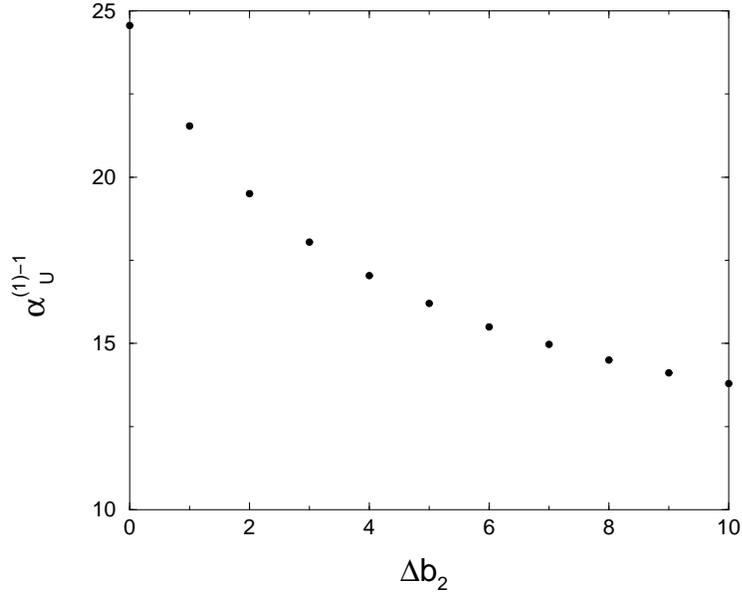,width=0.6\linewidth}
\caption{Plot of $\alpha_U^{(1)\,-1}$ as a function of $\Delta b_2$ under the
same conditions as in Fig.~\ref{fig:b1-b2}.}     
\label{fig:b2-agut}
\end{figure}

\begin{figure}[ht]
\centering
\epsfig{file=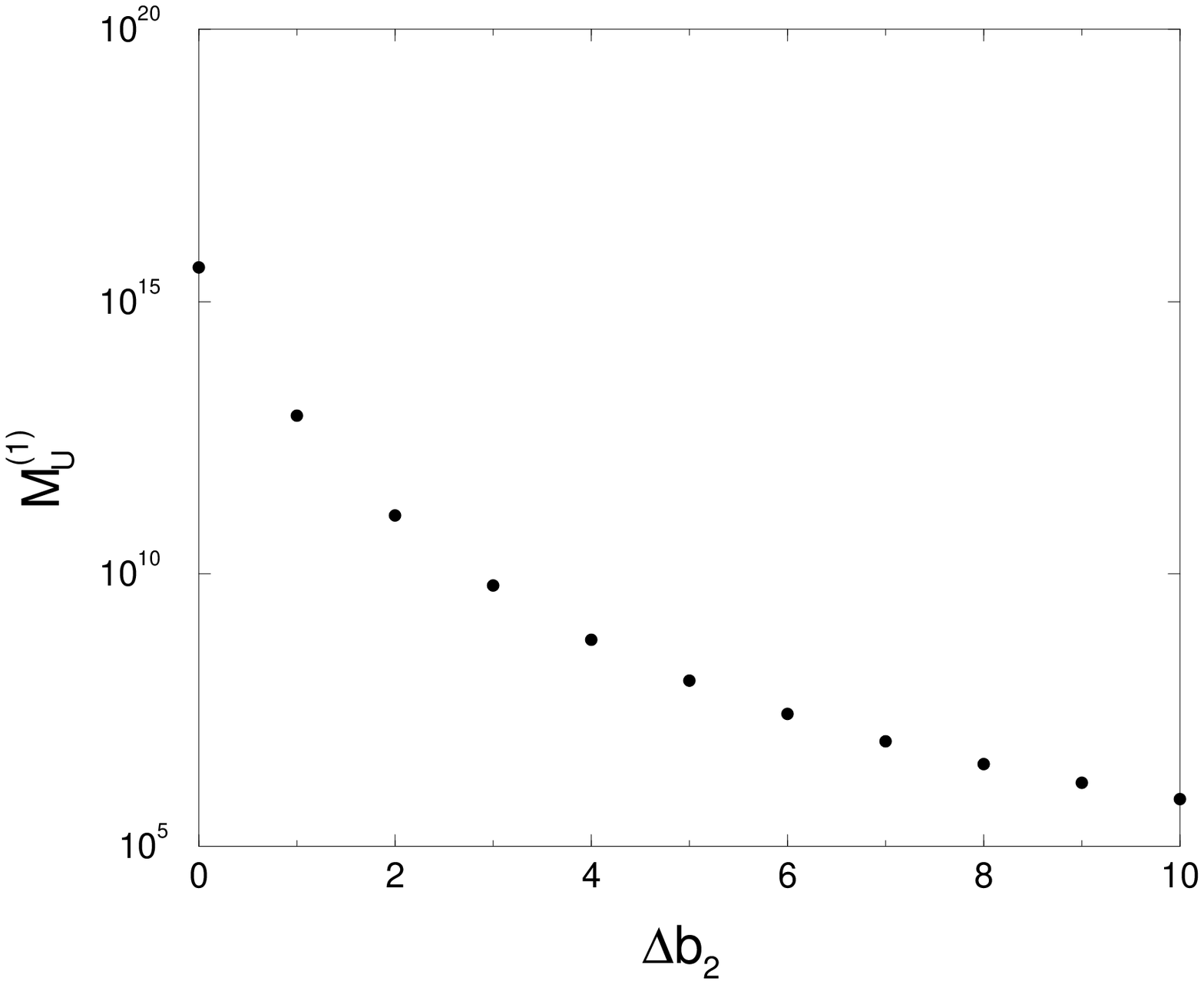,width=0.6\linewidth}
\caption{Plot of $M_U^{(1)}$ as a function of $\Delta b_2$ under the
same conditions as in Fig.~\ref{fig:b1-b2}.}
\label{fig:b2-mgut}
\end{figure}

Let us consider for the moment $\Delta b_3=0$, $\Delta b_2=0,1,\dots$, any
value of $\Delta b_1$ and fix $\alpha_3^{(1)}(M_Z)$ to the preferred value 
given in Eq.~(\ref{desired}). Then we can determine $\Delta b_1$ as a function
of $\Delta b_2$ which is shown in Fig.~\ref{fig:b1-b2}.
The corresponding values predicted for $\alpha_U^{(1)}$ are plotted in 
Fig.~\ref{fig:b2-agut}, which shows that all these solutions are well in the
perturbative regime, while the values of $M_U^{(1)}$ are plotted in 
Fig.~\ref{fig:b2-mgut}.
Notice first of all that the dots in Fig.~\ref{fig:b1-b2} do not 
necessarily correspond
to realistic models since, unlike $\Delta b_2$ which is being considered as
a positive integer, $\Delta b_1$ is unconstrained in Fig.~\ref{fig:b1-b2}. 
So the best we can do is
to try to approach the dots in Fig.~\ref{fig:b1-b2} in particular models. 
Second
of all, two-loop beta-coefficients do not have a direct dependence
on $\Delta b_i$ and so 
two-loop corrections are model dependent and must be computed in particular
models, as we will do in the following.

Before going to two-loop calculation in particular models we will discuss the
case where $\Delta b_3>0$. To cover the whole space $(\Delta b_1,\Delta b_2,
\Delta b_3)$ we can proceed as follows.
\begin{itemize}
\item
If $\Delta b_3 \leq \min(\Delta b_1,\Delta b_2)$, then we can make the global
shift $\Delta b_i\rightarrow \Delta b_i-\Delta b_3$ which shows that
the corresponding model
belongs to the class, in the sense of Eq.~(\ref{class}), of one
of the already considered models with $\Delta b_3=0$. The new model will have,
after two-loop corrections,
\begin{eqnarray}
\label{deltab3n0}
\alpha_U^{(\Delta b_3>0)}&>&\alpha_U^{(\Delta b_3=0)}\nonumber\\
M_U^{(\Delta b_3>0)}&<&M_U^{(\Delta b_3=0)}\nonumber\\
\alpha_3^{(\Delta b_3>0)}(M_Z)&\simeq &\alpha_3^{(\Delta b_3=0)}(M_Z)
\end{eqnarray}
so that it will not add anything new on the analysis of $\Delta b_3=0$ models.
\item
If $\Delta b_3 > \min(\Delta b_1,\Delta b_2)$ then:
\begin{itemize}
\item
If $\Delta b_2<\Delta b_3$ then all values $\Delta b_1\geq 0$ are to be 
considered.
\item
If $\Delta b_2\geq\Delta b_3$ then only values $\Delta b_1<\Delta b_3$ should
be considered. 
\end{itemize}

We have analyzed all these cases and found that, provided 
$M_U<M_{P\ell}$~\footnote{Of course for scales beyond $M_{P\ell}$ 
we should not trust the results of our field theory approach.}, 
there exist no model with $\Delta b_3\neq 0$ and  
$\alpha_3^{(1)}(M_Z)\simeq 0.105$.  
\end{itemize}

This closes the discussion on the $\Delta b_3\neq 0$ case and simplifies our
subsequent discussion on specific models.

\subsubsection{Canonical hypercharge models}

We want now to present the prediction for a very general class of models
containing an arbitrary amount of extra colorless matter. Models
containing color fields can be easily obtained from the latter
ones by looking for cases with beta-coefficients provided by the
shift $\Delta b_i\rightarrow\Delta b_i+\Delta b_3$ in agreement
with Eq.~(\ref{deltab3n0}) and the earlier comments.
We will choose
all those fields contained in $SU(5)$ representations $R_i$, such that
$d(R_i)\leq 24$, with arbitrary number as:
\begin{equation}
\label{contcan}
N_1 (\mathbf{1},\pm 1)+N_2 (\mathbf{2},\pm 1/2)+
N_3 (\mathbf{3},\pm 1)+n_3 (\mathbf{3},0)
\end{equation}
where the two labels correspond to the $SU(2)\times U(1)_Y$ quantum numbers,
and we are summing over the $\pm$ hypercharges to cancel
anomalies in a trivial way. 
\begin{figure}[ht]
\centering
\epsfig{file=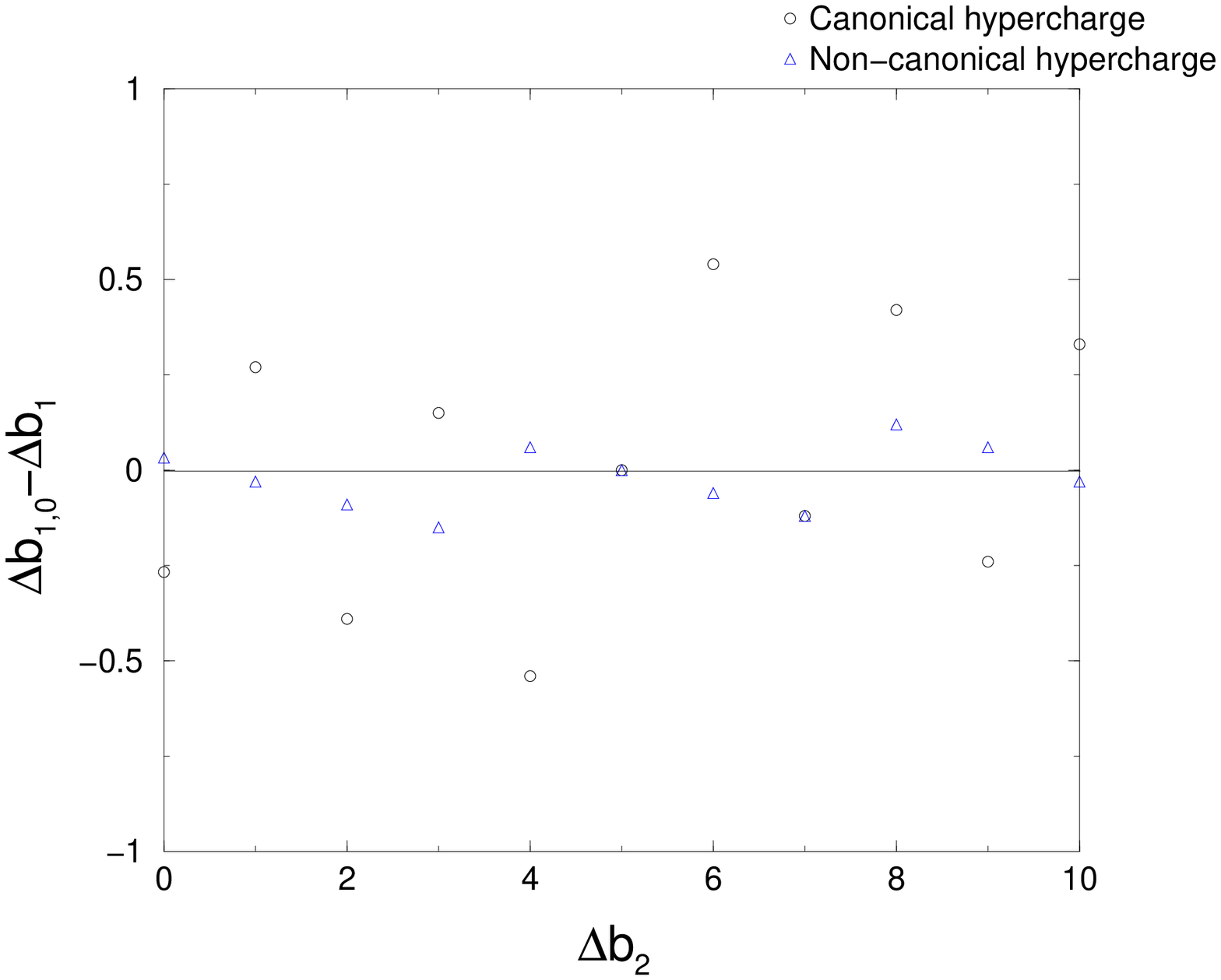,width=0.6\linewidth}
\caption{Nearest models, with respect to the dots in Fig.~\ref{fig:b1-b2} 
$(\Delta b_{1,0},\Delta b_2)$, 
which belong to the set defined by Eq.~(\ref{contcan}) (open circles)
and Eq.~(\ref{contst}) (open triangles).}
\label{fig:diffb1-b2}
\end{figure}

The contribution of these fields to the one-loop $\Delta b_i$, and two-loop
$\Delta b_{ij}$, beta-coefficients is given by:
\begin{eqnarray}
\label{delta1}
\Delta b_1 & = & \frac{3}{5}\,\left(2\, N_1+N_2+6\, N_3\right)\nonumber\\
\Delta b_2 & = & N_2+4\, N_3 +2\, n_3
\end{eqnarray}
and
\begin{eqnarray}
  \label{bij}
\Delta b_{11}& =& \frac{1}{25}\,\left(72\,N_1+9\,N_2+216\,N_3\right)
\nonumber\\
\Delta b_{12}&=&\frac{1}{5}\,\left(9\,N_2+144\,N_3\right)\nonumber\\
\Delta b_{21}&=&\frac{1}{5}\,\left(3\,N_2+48\,N_3\right)\nonumber\\
\Delta b_{22}&=&7\,N_2+48\,N_3+24\,n_3
\end{eqnarray}
Of course, the choice of the representations in (\ref{contcan}) is
arbitrary and should be considered for illustrative purposes. However,
higher dimensional representations would not alter
the general analysis since they contribute to 
$\Delta b_{2}$ with higher values, and their net
effect could always be replaced by a set of lower dimensional representations.
In particular the colorless representations with integer electric charge can be
written in terms of $SU(2)\times U(1)$ quantum numbers as:
$(\mathbf{2J+1},\pm Y)$ with $J=0,\,1/2,\,1,\,3/2,\,2,\dots$ and $Y$ an integer
(half-integer) for $J$ integer (half-integer). Their contribution to the
beta-coefficients is increasing with the dimension as $\Delta
b_2(\mathbf{1},\mathbf{2},\mathbf{3},\mathbf{4},\mathbf{5},\dots)=
(0,1/2,2,5,10,\dots)$.

The nearest models in the $(\Delta b_1,\Delta b_2)$-plane, with respect to the dots in 
Fig.~\ref{fig:b1-b2} with ordinate denoted by $\Delta b_{1,0}$, 
which belong to the set of models defined in Eq.~(\ref{contcan}) are shown 
in Fig.~\ref{fig:diffb1-b2}
(see open circles) for values of $\Delta b_2 \leq 10$. The particular values 
of the parameters
$N_1,N_2,N_3,n_3$ for the corresponding models are exhibited in 
Table~\ref{tab:cano} where the predictions
of $\alpha_3^{(1)}(M_Z)$ and $\alpha_3(M_Z)$ (two-loop calculation) are also 
shown. 
When two models are shown (which is the case for 
$\Delta b_2\leq 4$) the nearest model corresponds to that on the left column, 
and the 
next-to-nearest model to that on the right column. We can see that models 
with $\Delta b_1>\Delta b_{1,0}$, 
which are located in the lower half-plane of Fig.~\ref{fig:diffb1-b2}
($\Delta b_1<\Delta b_{1,0}$, located in the upper half-plane of 
Fig.~\ref{fig:diffb1-b2}) have 
$\alpha_3^{(1)}(M_Z)>0.105$ ($\alpha_3^{(1)}(M_Z)<0.105$) and the two-loop 
calculation gives values of
$\alpha_3(M_Z)$ which are larger (smaller) than the experimental value 
(\ref{strong}). In fact for the values of $\Delta b_2$ for which two models
are presented, one of them predicts a value of $\alpha_3(M_Z)$ smaller and
the other larger than the experimental value (\ref{strong}). 

Two-loop corrections are computed using the beta-coefficients (\ref{bij}).
Extra matter is considered at the TeV scale and we have assumed absolute
degeneracy in the supermultiplets and no threshold effect.
\begin{table}[ht]
\begin{center}
\begin{tabular}{|c|c|c|c|c|c|c|c|}
\hline
$\Delta b_1$&$\Delta b_2$&$N_1$&$N_2$&$N_3$&$n_3$&$\alpha_3^{(1)}(M_Z)$&
$\alpha_3(M_Z)$\\
\hline
0,\ 6/5&0&0,1&0,0&0,0&0,0&0.117,\ 0.083&0.131,\ 0.087\\
3,\ 9/5&1&2,1&1,1&0,0&0,0&0.099,\ 0.148&0.106,\ 0.181\\
24/5,\ 6&2&4,5&0,0&0,0&1,1&0.117,\ 0.091&0.134,\ 0.097\\
39/5,\ 33/5&3&6,5&1,1&0,0&1,1&0.103,\ 0.137&0.114,\ 0.168\\
48/5,\ 54/5&4&5,6&0,0&1,1&0,0&0.117,\ 0.096&0.135,\ 0.104\\
63/5&5&7&1&1&0&0.106&0.118\\
78/5&6&10&0&1&1&0.099&0.110\\
87/5&7&11&1&1&1&0.108&0.122\\
102/5&8&1&0&2&0&0.101&0.113\\
111/5&9&12&1&2&0&0.109&0.124\\
126/5&10&15&0&2&1&0.103&0.117\\
\hline
\end{tabular}   
\caption{Values of $\alpha_3 (M_Z)$ for models with canonical
hypercharge.}
\label{tab:cano}
\end{center}
\end{table}
From Table~\ref{tab:cano} we see that there are models which
belong to the same equivalence class, in the sense of Eq.~(\ref{class}). 
For instance, models 
on the left column with $\Delta b_2=0,2,4$ belong to the MSSM equivalence 
class and could have been found
just by exploring the corresponding equation (\ref{classMSSM}). We can also
see that in order to corner values of $\alpha_3(M_Z)$ inside the experimental
range (\ref{strong}) we have to go to $\Delta b_2=5$ 
$(N_1=7,\, N_2=N_3=1)$ and this model predicts
values of the unification scale and gauge coupling given by: 
$M_U=5\times 10^{8}$ GeV and $\alpha_U=1/15.4$.

Let us finally stress that all presented models have $\Delta b_3=0$.
Following our general comments at the beginning of this section, models
with $\Delta b_3>0$ can be found, which belong to the same class as 
models with $\Delta b_3=0$, and with unification scales and couplings as
given by Eq.~(\ref{deltab3n0}). In particular they have smaller unification
scale than $\Delta b_3=0$ models. Therefore the requirement of gauge
coupling unification has led us to a general 
upper bound on the unification scale
as: $M_U\lesssim 5\times 10^{8}$ GeV.

\subsubsection{Non-canonical hypercharge models}
Of course with non-canonical hypercharge assignment there is much more
freedom in the choice of extra matter. For the sake of illustration we will
consider a particular $SU(4)_c\times SU(2)_L\times SU(2)_R$ string 
model~\cite{ALT} that breaks to
$SU(3)_c\times SU(2)_L\times U(1)_Y$ at the scale $M_U$. If the latter
scale coincides with the string scale $M_{st}$ the gauge couplings should unify
at $M_U$ and we can constrain its value from the unification condition, 
similarly to what we have done in the previous section for general models with
canonical hypercharge assignment. The $SU(3)_c\times SU(2)_L\times U(1)_Y$
quantum numbers of the extra matter is:
\begin{eqnarray}
\label{contst}
&&N_2\,(\mathbf{1},\mathbf{2},\pm 1/2)+N'\,(\mathbf{1},\mathbf{1},\pm 1/2)
+2\,N_L\,(\mathbf{1},\mathbf{2},0)\nonumber \\
&+&
N_3\,(\mathbf{3},\mathbf{1},\pm 1/3)+N'_3\,(\mathbf{3},\mathbf{1},\pm 1/6)
+N_{31}\,(\mathbf{3},\mathbf{1},\pm 2/3)
\end{eqnarray}
where we assume we are summing over $\pm$ non-zero hypercharges so the
corresponding labels count the number of pairs. On the other hand
we can see there can be colorless matter with fractional electric
charge, which is a general feature in many string models.

The field content of (\ref{contst}) contribute to the one-loop and two-loop
beta functions, $\Delta b_i$ and $\Delta b_{ij}$, respectively. However,
following the general analysis at the beginning of this section, in the
search of models and their unification scales $M_U$, it is enough to consider
the cases where $\Delta b_3=0$ and $\Delta b_{3i}=\Delta b_{i3}=0$,
which amounts to take into account only colorless states, those on the first
row of Eq.~(\ref{contst}). Their contribution to the beta-coefficients can be
written as:
\begin{eqnarray}
\label{Deltap}
\Delta b_1 &=& \frac{3}{5}\, \left( N_2+\frac{N'}{2}\right)\nonumber\\
\Delta b_2 &=& N_2+N_L
\end{eqnarray}
and
\begin{eqnarray}
\label{bijp}
\Delta b_{11}& =& \frac{1}{25}\,\left(72\,N_1+\frac{9}{2}\,N'\right)
\nonumber\\
\Delta b_{12}&=&\frac{9}{10}\,N_2\nonumber\\
\Delta b_{21}&=&\frac{3}{5}\,N_2\nonumber\\
\Delta b_{22}&=&7\,N_2+7\,N_L
\end{eqnarray}
The nearest models in the $(\Delta b_1,\Delta b_2)$-plane, with respect to 
the dots in Fig.~\ref{fig:b1-b2}, 
which belong to the set of models defined in Eq.~(\ref{contst}) are shown 
in Fig.~\ref{fig:diffb1-b2}
(see open triangles) for values of $\Delta b_2 \leq 10$. The particular value 
of the parameters
$N_2,N',N_L$ for the corresponding models is exhibited in 
Table~\ref{tab:non-cano} where the predictions
of $\alpha_3^{(1)}(M_Z)$ and $\alpha_3(M_Z)$ (two-loop calculation) are also 
shown. We can see (as in the case of canonical hypercharge) that models 
with $\Delta b_1>\Delta b_{1,0}$, 
which are located in the lower half-plane of Fig.~\ref{fig:diffb1-b2}
($\Delta b_1<\Delta b_{1,0}$, located in the upper half-plane of 
Fig.~\ref{fig:diffb1-b2}) have 
$\alpha_3^{(1)}(M_Z)>0.105$ ($\alpha_3^{(1)}(M_Z)<0.105$) and the two-loop 
calculation gives values of
$\alpha_3(M_Z)$ which are larger (smaller) than the experimental value 
(\ref{strong}). 
Two-loop corrections are computed using the beta-coefficients (\ref{bijp}).
Again extra matter is considered at the TeV scale and we have assumed absolute
degeneracy in the supermultiplets and no threshold effect.
\begin{table}[ht]
\begin{center}
\begin{tabular}{|c|c|c|c|c|c|c|}
\hline
$\Delta b_1$&$\Delta b_2$&$N_2$&$N_L$&$N^{\prime}$&$\alpha_3^{(1)}(M_Z)$&
$\alpha_3(M_Z)$\\
\hline
3/10&0&0&0&1&0.105&0.114\\
27/10&1&1&0&7&0.107&0.116\\
51/10&2&1&1&15&0.108&0.119\\
36/5&3&2&1&20&0.117&0.131\\
51/5&4&2&2&30&0.105&0.114\\
63/5&5&4&1&34&0.106&0.116\\
15&6&4&2&41&0.107&0.117\\
87/5&7&5&2&48&0.108&0.118\\
201/10&8&6&2&55&0.105&0.115\\
45/2&9&6&3&63&0.105&0.116\\
249/10&10&7&3&69&0.106&0.117\\
\hline
\end{tabular}
\caption{Values of $\alpha_3 (M_Z)$ for models with non-canonical hypercharge.}
\label{tab:non-cano}
\end{center}
\end{table}

We can see that the prediction of $\alpha_3(M_Z)$ for the models from 
Table~\ref{tab:non-cano} is much closer to the experimental range 
(\ref{strong}) than for those with canonical hypercharge assignment from
Table~\ref{tab:cano}. In particular the model with $\Delta b_2=0$ predicts
a value of $\alpha_3(M_Z)$ which is 2.5$\sigma$, and the model with
$\Delta b_2=1$ is only 1.5$\sigma$. 
However the first model where $\alpha_3(M_Z)$
is inside the experimental range corresponds to 
$\Delta b_2=2$, which unifies at $M_U=5\times 10^{11}$ GeV with 
$\alpha_U=1/18.5$.

The class of models defined by Eq.~(\ref{contst}) has been
recently analyzed by the authors of Ref.~\cite{LT}. 
The values for the matter content
$(N_2,N',N_L)$ they obtain for acceptable models do not coincide with 
those appearing in Table~\ref{tab:non-cano}. The reason is that
in Ref.~\cite{LT} only the one-loop RGE analysis has been performed
while we are including two-loop corrections (and MSSM weak scale
threshold effects), which are sizable as stated in Eq.~(\ref{inc2}).

To conclude, in the class of string models defined by the extra particle
content (\ref{contst}), strong coupling unification implies that
$M_U\lesssim 5\times 10^{11}$ GeV.

\section{Power-law unification}
\label{power-law}

Another possibility, which has been recently stressed by 
Dienes, Dudas and Gherghetta~\cite{dienes}, is
based on a $4+p$ dimensional theory with a (common) compactification radius
for the extra $p$ dimensions $M_c^{-1}$, with 
$M_c\ll M_U$, which leads to much lower values of the unification scale
$M_U$.
Under these conditions, the $p$ extra dimensions open up at the scale $M_c$ and
the gauge coupling renormalization receive the contribution of the towers
of Kaluza-Klein (KK) excitations which make them to run with a power-low
behaviour~\cite{vt}. 
Under certain circumstances, that we will analyze in full generality
next, the gauge couplings unify shortly after $M_c$. In the case where there is
a hierarchy among the radii of $p$ dimensions, the theory for scales greater
than $1/R_5$ behaves as five dimensional, evolution of the gauge couplings is
accelerated and they may unify at a scale $\lesssim 1/R_6$. For this reason
we will consider, for the computational sake, the case of a 5D theory, 
with $M_c$ the inverse radius of the fifth dimension, although our results 
are completely general. 

We will consider the 5D theory compactified on the fifth dimension orbifold
$S^1/\mathbb{Z}_2$~\cite{peskin,alex}. 
Half of the KK modes is projected away by the action of
$\mathbb{Z}_2$, which means that for zero-modes the theory has $N=1$
supersymmetry, while non-zero modes keep their $N=2$ nature and utilize the
odd-modes to reconstruct entire KK-towers, whose number is therefore divided
by two. A full explanation of this kind of models can be found in 
Refs.~\cite{alex}. The MSSM is made up of zero-modes of fields living in the
5D bulk and fields that can live in the 4D boundary, at localized points of
the bulk. (Fixed points in the language of the heterotic string.) The fields
that live on the 4D boundary are arranged in $N=1$ supermultiplets and those
living in the bulk in $N=2$ supermultiplets. 

Vector fields in
the bulk are in $N=2$ vector supermultiplets. For non-zero (massive) KK-modes
they contain a massive vector field, a real scalar and a Dirac fermion. For
zero (massless) KK-modes, only the $N=1$ vector supermultiplet survive the
$\mathbb{Z}_2$ projection. Chiral fields in the bulk belong to $N=2$
hypermultiplets. They contain two complex scalars and a Dirac fermion.
For the zero-modes only one complex scalar and a particular chiral projection
of the Dirac fermion survive after $\mathbb{Z}_2$ projection. 
This is the case of the Higgs and matter sector $(H_1,H_2,Q,U,D,L,E)$. 
Whenever they live in the bulk, they should constitute an independent 
hypermultiplet which, after $\mathbb{Z}_2$ projection leads to the
supermultiplet structure of the MSSM~\footnote{This is in contrast with the
particle content of Refs.~\cite{dienes,carone}, 
where both $H_1$ and $H_2$ of the
MSSM belong to the same hypermultiplet. In this case, after
$\mathbb{Z}_2$ projection one of the MSSM is projected out and cannot give
mass to the corresponding MSSM fermion.}.

The generic scenario is the following. We have a 4D model valid up
to the compactification scale $M_c$. Its particle content 
consists on (half of) the zero-modes of fields living in the bulk plus
fields living on the boundary. The 4D fields are arranged in $N=1$ 
supermultiplets and the gauge coupling evolution is logarithmic. The 4D model
can be any of the models previously considered. Given the particular model,
characterized by beta-coefficients $b_i$ and a particular value of
$\alpha_3^{(1)}(M_Z)$, from the 4D formula (\ref{alphas1}), when the extra
dimension opens up at the scale $M_c$ the states that contribute to the
gauge coupling renormalization for scales beyond $M_c$ are: 
\begin{enumerate}
\item
Fields living on the boundary, that contribute logarithmically to the 
renormalization of the gauge couplings.
\item
Fields living in the bulk which constitute
towers of $N=2$ supermultiplets, with beta functions $\widetilde{b}_i$, that
contribute with a power-law to the gauge coupling renormalization. 
\end{enumerate}
The one-loop evolution is then given by:
\begin{equation}
\label{1LKK}
\alpha^{-1}_i(\Lambda)=\alpha^{-1}_i(\mu)-\frac{b_i}{2\pi}\log
\frac{\Lambda}{\mu} 
-\frac{\widetilde{b}_i}{2\pi}\left( \frac{\Lambda}{M_c}-1-\log\frac{
\Lambda}{M_c}\right)
\end{equation}

From Eq.~(\ref{1LKK}) we can deduce analytical equations for $M_U^{(1)}$,
$\alpha_U^{(1)}$ and $\alpha_3^{(1)}(M_c)$ by imposing unification and
neglecting the logarithmic running between $M_c$ and $M_U$. We obtain
equations similar to those obtained for the logarithmic running in 4D
(\ref{MU1}), (\ref{aU1}) and (\ref{alphas1}):
\begin{equation}
\label{MU1KK}
M_U^{(1)}=\alpha_{12}^{-1}(M_c)\,\frac{2\pi}
{\widetilde{b}_1-\widetilde{b}_2}\ M_c
=\frac{b_1-b_2}
{\widetilde{b}_1-\widetilde{b}_2}\ M_c\ \log\frac{M_U^{(1)\,{\rm 4D}}}{M_c}
\end{equation}
where $M_U^{(1)\,{\rm 4D}}$ is the 4D prediction of $M_U$ as given by 
(\ref{MU1}),
\begin{equation}
\label{aU1KK}
\alpha_U^{(1)\, -1}=\alpha_1^{-1}(M_c)-\alpha_{12}^{-1}(M_c)\ 
\frac{\widetilde{b}_1}
{\widetilde{b}_1-\widetilde{b}_2}=\alpha_2^{-1}(M_c)-
\alpha_{12}^{-1}(M_c)\ 
\frac{\widetilde{b}_2}{\widetilde{b}_1-\widetilde{b}_2}
\end{equation}
\begin{equation}
\label{alphas1KK}
\alpha_3^{(1)\, -1}(M_c)=\alpha_2^{-1}(M_c)-\alpha_{12}^{-1}(M_c)\ 
\frac{\widetilde{b}_2-\widetilde{b}_3}
{\widetilde{b}_1-\widetilde{b}_2}=
\alpha_1^{-1}(M_c)-\alpha_{12}^{-1}(M_c)\ 
\frac{\widetilde{b}_1-\widetilde{b}_3}
{\widetilde{b}_1-\widetilde{b}_2}
\end{equation}
We can match the prediction of the theory for $\alpha_3^{(1)}(M_c)$
running with a power law from $M_U$, as given by (\ref{alphas1KK}), with
that of the 4D theory running logarithmically from $M_Z$, as given by
(\ref{alphas1}) with $\mu=M_c$. Then the 5D theory predicts at one-loop 
the same value 
of $\alpha_3^{(1)}(M_Z)$ as the 4D theory provided that the conditions
\begin{equation}
\label{ecKK}
\frac{b_1-b_3}{b_1-b_2}=
\frac{\widetilde{b}_1-\widetilde{b}_3}
{\widetilde{b}_1-\widetilde{b}_2},\quad
 \frac{b_2-b_3}{b_1-b_2}=
\frac{\widetilde{b}_2-\widetilde{b}_3}
{\widetilde{b}_1-\widetilde{b}_2}
\end{equation}
hold, whose general solution is:
\begin{equation}
\label{classKK}
\epsilon^{ijk}\left(b_i-b_j\right)\,\widetilde{b}_k=0
\end{equation}
Notice that the equation (\ref{classKK}) is similar to the class condition
for 4D models (\ref{class}), which shows that the strong coupling unification
condition at one-loop is independent on the dimensionality.

We will now consider the case of the MSSM for two reasons.
On the one hand simplicity, since our aim
is to prove that the MSSM can solve the strong unification problem in the
presence of large extra dimensions, and find an upper bound on the value of 
the compactification and unification scales. The second reason is that models
considered in section \ref{logarithmic} would lead to a lower value of the
unification scale than that provided by the MSSM and the study of those models
would not add anything new concerning the former upper bound.

In the case of the MSSM the unification condition (\ref{classKK}) reads as:
\begin{equation}
\label{clMSSMKK}
5\,\widetilde{b}_1-12\,\widetilde{b}_2 +7\,\widetilde{b}_3=0\, 
\end{equation}
where now $\widetilde{b}_i$ are the $N=2$ beta-coefficients, given by:
\begin{equation}
\widetilde{b}_i=-2 \left(C_2(G_i)-T(R)\right)
\label{N=2}
\end{equation}
and $R$ are the representations which the hypermultiplets belong to. Assuming
a general $N=2$ superfield content, with $N_X$ hypermultiplets $X$ 
$(X=Q,U,D,L,E,H_1,H_2)$ and $\delta_{G}$ vector multiplets $(G=SU(3),SU(2),
U(1))$ living in the bulk $(\delta_G=0,1)$, the condition (\ref{clMSSMKK}) can
be written as:
\begin{equation}
\label{claseKK}
2\,N_E+5\,N_U-7\,N_Q+3\,\left(N_D-N_L\right)=14\,\delta_{SU(3)}
-16\,\delta_{SU(2)}+3\,\left(N_{H_1}+N_{H_2}\right)
\end{equation}

However, not all values of the parameters are attainable in 
Eq.~(\ref{claseKK}), since we must require that hypermultiplets in the bulk
experience gauge interactions in the $N=2$ theory. Therefore, on
general grounds we can split Eq.~(\ref{claseKK}) into three different possible
cases, depending on the possible values of $\delta_{G_i}$.

\begin{enumerate}
\item
The case where the $SU(3)\times SU(2)\times U(1)$ gauge bosons are in the bulk.
In this case $\delta_{SU(3)}=\delta_{SU(2)}=1$ and Eq.~(\ref{claseKK}) reads as
\begin{equation}
\label{claseKK1}
2\,N_E+5\,N_U-7\,N_Q+3\,\left(N_D-N_L\right)-3\,\left(N_{H_1}+N_{H_2}\right)
+2=0
\end{equation}
\item
The case where only the $SU(3)\times U(1)$ gauge bosons are in the bulk.
In this case $\delta_{SU(3)}=1,\, \delta_{SU(2)}=0$, $N_Q=N_L=N_{H_i}=0$
(no $SU(2)$-doublets in the bulk), and Eq.~(\ref{claseKK}) 
reads as
\begin{equation}
\label{claseKK2}
2\,N_E+5\,N_U+3\,N_D=14
\end{equation}
\item
The case where only the $SU(2)\times U(1)$ gauge bosons are in the bulk.
In this case $\delta_{SU(3)}=0,\, \delta_{SU(2)}=1$, $N_U=N_Q=N_D=0$ 
(no $SU(3)$-triplets in the bulk), and Eq.~(\ref{claseKK}) reads as
\begin{equation}
\label{claseKK3}
2\,N_E+16=3\,\left(N_L+N_{H_1}+N_{H_2} \right)
\end{equation}
\end{enumerate}

Notice that the $U(1)$ gauge multiplet should be situated 
in the bulk, if hypermultiplets with hypercharge live in the
bulk, since they should experience the corresponding gauge interactions. 
In the particular case of no hypermultiplets in the bulk, the
$U(1)$ gauge multiplet can be either in the bulk or on the boundary.
This leads to
the trivial solution to Eq.~(\ref{claseKK}) where all matter,
except the $U(1)$ gauge boson, is at the boundary. In this case
the beta-coefficients $\widetilde{b}_i=0$,  the evolution
beyond $M_c$ is logarithmic and the unification is like in the
4D MSSM, Eqs.~(\ref{predMSSM}). Still the $U(1)$ in the bulk can
be used to break supersymmetry by the Scherk-Schwarz mechanism
by giving a mass to the $\widetilde{B}$ gaugino, with the
subsequent propagation of supersymmetry breaking to the boundary
by radiative corrections~\cite{alex}.

Any matter content in the bulk consistent with Eq.~(\ref{claseKK})
will then provide a model where the one-loop prediction of
$\alpha_3(M_Z)$ is as in the MSSM but with $M_U$ and $\alpha_U$
differently determined, Eqs.~(\ref{MU1KK}) and (\ref{aU1KK}).
Particular cases satisfying Eqs.~(\ref{claseKK1})-(\ref{claseKK3}) 
have recently been proposed in the literature: 

\begin{itemize}
\item
The case where
the gauge and Higgs sector, along with two $E$ hypermultiplets,
is in the bulk,
$(\delta_{SU(3)}=\delta_{SU(2)}=N_{H_1}=N_{H_2}=1, N_E=2)$, has
been discussed by Kakushadze~\cite{zurab}. It belongs to case 1 before and
satisfies Eq.~(\ref{claseKK1}). Other possibilities are obvious from
Eq.~(\ref{claseKK1}), apart from those corresponding to complete $SU(5)$
representations: $N_E=N_U=N_Q$ and/or $N_D=N_L$.

\item
The case $\delta_{SU(3)}=1,\, \delta_{SU(2)}=0$, $N_U=N_D=1$, $N_E=3$ has 
been discussed recently by Carone~\cite{carone}. 
It satisfies Eq.~(\ref{claseKK2}) and belongs to case 2 before.
Other trivial possibilities are, e.g.: 
a) $N_U=N_E=2$, the other $N_i$ equal to zero;
b) $N_D=2$, $N_E=4$, the other $N_i$ equal to zero.

\item
The case $\delta_{SU(3)}=0,\, \delta_{SU(2)}=1$, $N_E=N_{H_1}=1$,
$N_L=5$ has also been
discussed in Ref.~\cite{carone}. It satisfies Eq.~(\ref{claseKK3})
and belongs to case 3 before.
There are of course other trivial possibilities satisfying Eq.~(\ref{claseKK3})
as, e.g.: a) $N_E=N_{H_i}=1$, $N_L=4$; b) $N_E=4$, $N_{H_i}=1$, $N_L=6$.
\end{itemize}

It is clear that a
great variety of models satisfying Eq.~(\ref{claseKK}) can be
discussed and proposed, leading to very similar values of the
unification scale and coupling at the one-loop level. In all
cases the hypermultiplets in (\ref{claseKK}) can be
considered either as extra matter (with respect to the MSSM) 
which decouples at $M_c$~\footnote{This can be achieved by means
of a supersymmetric mass term in the superpotential~\cite{zurab}.},
or as part of the MSSM~\footnote{In this case the problem of giving a mass to
the corresponding surviving zero-mode fermion by the electroweak breaking 
mechanism has to be resolved explicitly in each case~\cite{alex}.}.

\begin{figure}[ht]
\centering
\epsfig{file=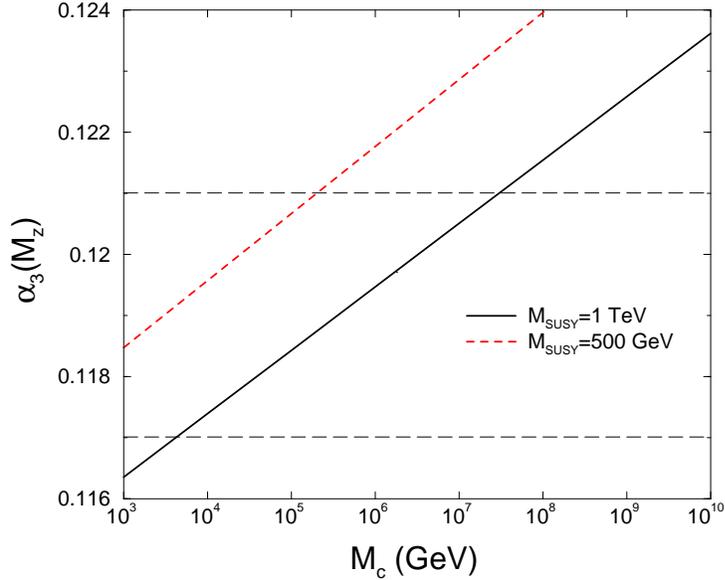,width=0.6\linewidth}
\caption{Value of $\alpha_3 (M_Z)$ versus $M_c$. The two horizontal lines
represent the experimental 1$\sigma$ band.}
\label{fig:alpha}
\end{figure}

To make a more accurate prediction for this class of models we
will include, as in Ref.~\cite{graham}, two-loop effects between the
weak scale and $M_c$, the $\overline{\rm MS}$ to $\overline{\rm
DR}$ conversion factors and the MSSM supersymmetric threshold
effects as in section~\ref{logarithmic}. Then the RGE
predictions for the gauge couplings can be written as:
\begin{eqnarray}
\label{RGEKK}
\alpha_i^{-1}(M_Z)&=&\alpha_U^{-1}+\frac{b_i}{2\pi}\log\frac{M_U}{M_Z}+
\frac{1}{4\pi}\sum_{j=1}^{3}\frac{b_{ij}}{b_j}\log\frac{\alpha_j(M_U)}
{\alpha_j(M_Z)}+\Delta_i\nonumber \\
&+&\frac{\widetilde{b}_i}{2\pi}\left(\frac{M_U}{M_c}-1-\log\frac{M_U}{M_c}
\right)
\end{eqnarray}

The numerical results are shown in Fig.~\ref{fig:alpha} where the prediction 
of $\alpha_3(M_Z)$ is plotted versus $M_c$ for two particular values of the
common supersymmetry breaking mass $M_{\rm SUSY}$ (500 GeV and 1 TeV). Then
from the experimental band for $\alpha_3(M_Z)$ we obtain an upper bound for
$M_c$ as $M_c\lesssim 2\times 10^7$ GeV for $M_{\rm SUSY}$=1 TeV and
$M_c\lesssim 1.5\times 10^5$ GeV for $M_{\rm SUSY}$=500 GeV. The corresponding
values of $M_U$ can be obtained easily from (\ref{MU1KK}). There is a mild
model dependence from the term $b_{12}/\widetilde{b}_{12}$, although this
term is usually $\mathcal{O}(1)$. To fix the ideas, and as an illustration, in
the model with $(\delta_{SU(3)}=\delta_{SU(2)}=N_{H_1}=N_{H_2}=1, N_E=2)$ of
Ref.~\cite{zurab}, $(b_1-b_2)/(\widetilde{b}_1-\widetilde{b}_2)=1$, and 
$M_U\lesssim 4.4\times 10^8$ GeV for $M_{\rm SUSY}$=1 TeV,
$M_U\lesssim 4\times 10^6$ GeV for $M_{\rm SUSY}$=500 GeV.

A word of caution has to be said about this result. In our RGE for the
evolution of the gauge couplings from $M_Z$ to $M_U$ we have considered 
two-loop effects below $M_c$ as well as supersymmetric threshold effects at 
the weak scale. However no two-loop effect beyond $M_c$ has been considered.
This issue has been studied in Ref.~\cite{zurab} where, using the fact that 
the gauge coupling is not renormalized beyond one-loop in the $N=2$ 
theory~\footnote{Remember that the $\mathbb{Z}_2$ projection spoils the $N=2$ 
supersymmetry for the zero-modes and that matter on the boundary is $N=1$ 
supersymmetric.}, it was estimated that the ratio of two-to-one loop
renormalization of the inverse gauge coupling goes as:
$$\Delta^{(2)}\,\alpha_i^{-1}=\mathcal{O}\left(\frac{M_c}{\Lambda}\right)
\Delta^{(1)}\,\alpha_i^{-1}$$ 
where $\Lambda$ is identified here as the
cutoff of the theory. This means that perturbation theory seems
to be trustable but nevertheless the (accelerated) unification predictions 
are not guaranteed to survive to higher loop corrections. A preliminary
calculation of two-loop corrections for scales beyond $M_c$ has shown that
the leading corrections, in particular those with bulk and boundary
fields mixed in the diagram, are very model dependent. In particular
they depend on the distribution of matter in bulk hypermultiplets and 
boundary chiral multiplets. A detailed calculation of these two-loop
corrections is outside the scope of this paper and will be postponed for 
future work. 

Finally, let us stress that models with extra dimensions at the TeV scale are
favoured by gauge coupling `accelerated' unification. In this case the extra
dimensions could be used to trigger supersymmetry breaking by the 
Scherk-Schwarz~\footnote{The Scherk-Schwarz mechanism has been
used to spontaneously break local supersymmetry in
string~\cite{SSstring} and $M$-theory~\cite{SSM} and always leads
to a gravitino mass of the order of the inverse compactification
radius which breaks supersymmetry. In this case we
would have $m_{3/2}\sim M_c$.}
mechanism~\cite{alex}, and lead to observable effects 
in high-energy colliders by the
production of Kaluza-Klein excitation of the MSSM 
particles~\cite{ignatios,KKprod}. 
In particular, if we fix in
Fig.~\ref{fig:alpha} $M_c=1$ TeV one obtains
$\alpha_3(M_Z)=0.1165$ for $M_{\rm SUSY}=1$ TeV and
$\alpha_3(M_Z)=0.1185$ for $M_{\rm SUSY}=0.5$ TeV, in rather
good agreement with precision data.

\section{Conclusions}
\label{conclusions}
In this paper we have analyzed the unification of $SU(3)\times SU(2)\times
U(1)$ gauge couplings in supersymmetric extensions of the Standard Model,
using the gauge couplings determination at $M_Z$ deduced from LEP and low 
energy precision measurements. We have considered two very general unification
scenarios:
\begin{description}
\item{\sf Logarithmic unification}

This is the case where unification is achieved in an
$N=1$ four-dimensional theory 
and, therefore, the evolution of gauge couplings is logarithmic. We impose
unification of general supersymmetric models and perturbativity, and obtain
the unification scale $M_U$, that we assume to be of the order of magnitude of
the compactification scale where new dimensions open up, $M_c$. 

\item{\sf Power-law unification}

In this case the extra dimensions open up at scales $M_c$ much lower than the
unification scale and, beyond $M_c$, gauge couplings run, as functions of the 
energy, with a power-law behaviour. Unification happens in an `accelerated'
way during the last stages of their evolution. The theory below $M_c$ is
$N=1$ supersymmetric and beyond $M_c$ the Kaluza-Klein states form multiplets
of $N=2$ supersymmetry. We consider, below $M_c$ the MSSM, and beyond it a
general class of $N=2$ models that provide the same one-loop strong coupling
at $M_Z$ as the MSSM. The experimental range for the gauge couplings provide a
corresponding band in $M_c$, and in $M_U$, with an upper bound, which 
implies in general large extra dimensions.
\end{description}

In the case of logarithmic unification we have examined the general set 
of models
with canonical hypercharge (such that colorless extra fields have integer
electric charge), and also a class of unifying string models with fractional 
electric charge for colorless states, with unification gauge group
$SU(4)\times SU(2)\times SU(2)$. The experimental upper bound on 
$\alpha_3(M_Z)$ translates into an upper bound on the unification scale which
is $\sim 10^{9}$ GeV for the canonical hypercharge models and  
$\sim 10^{12}$ GeV for string models. 

In the case of power-law unification we have examined
the general class of models which coincide with the MSSM in 4D, below the
compactification scale $M_c$, and that unify in $(4+p)$ dimensions with the
same one-loop value of $\alpha_3(M_Z)$ as the MSSM. Inclusion of two-loop
corrections below $M_c$ provides upper bounds on $M_c$ and $M_U$
($\sim 10^{7}$ GeV and $\sim 10^{9}$ GeV, respectively) from the unification
constrains. However, before drawing definite predictions in this case higher
loop corrections from Kaluza-Klein excitations should be precisely computed.

In either case, logarithmic or power-law unification, it is possible that the
size of the required extra dimension is $\sim$ TeV$^{-1}$. This fact makes it
possible that:
\begin{itemize}
\item
The extra dimension have observable effects in high-energy colliders. These
effects can either put lower bounds on the mass of the corresponding 
KK-excitations, in the case of negative searches, 
or enable their experimental direct discovery~\cite{ignatios,KKprod}.
\item
The extra dimensions be used, as a useful mechanism to break supersymmetry
by the Scherk-Schwarz compactification, creating a mass splitting between
members of the same supermultiplet. The main features of this mechanism have
been extensively studied~\cite{alex}.
\end{itemize}

All along this paper we have used GUT-like unification condition, i.e.
$\alpha_3(M_U)=\alpha_2(M_U)=k_1\alpha_Y(M_U)$ where $k_1=5/3$. This is the
case in grand unification and many string theories. However in some
string constructions other values of $k_1$ may appear that can help to solve 
in a different fashion the strong coupling problem. We wish to mention here
this possibility, which has been already stressed in the literature. Just
to give an illustrative example, in the case of $k_1=7/4$ the predictions of
the MSSM including the same corrections as those considered in 
section~\ref{logarithmic} are:
\begin{eqnarray}
\label{k1dif}
M_U&=&9.7\times 10^{15}\,\mathrm{GeV}\nonumber\\
\alpha_U&=&1/23.2\nonumber\\
\alpha_3(M_Z)&=& 0.116
\end{eqnarray}
where $\alpha_3(M_Z)$ is now in rough agreement with the experimental
data.

To conclude we have proved that the unification of gauge couplings, as deduced
from the electroweak precision data, requires in a general class of
models the presence of large extra dimensions 
(as large as $\mathrm{TeV}^{-1}$) and/or low unification scale.

\section*{Acknowledgements} 
We thank J.R. Espinosa and G. Kane for useful discussions. The work of AD 
was supported by the Spanish Education Office (MEC) under an \emph{FPI}
 scholarship.   
%

\end{document}